\documentclass[letterpaper, conference]{ieeeconf}  % Comment this line out if you need a4paper 

\IEEEoverridecommandlockouts  % This command is only needed if you want to use the \thanks command

\usepackage{amsthm} % For proof and use \qedhere to put box where you want

\newtheorem{assum}{Assumption}
\newtheorem{prop}{Proposition}
\newtheorem{defn}{Definition}
\newtheorem{problem}{Problem}
\newtheorem{subproblem}{Problem}

\usepackage{caption}
\usepackage{graphicx}
\usepackage{subcaption}
\graphicspath{{../Figures/}}

\usepackage{subcaption}
\usepackage{hyperref}
\usepackage{cite}
\usepackage{multirow}
\usepackage[export]{adjustbox}
\usepackage{amssymb} % For mathbb, latex symbols, checkmark
\usepackage{amsbsy}
\usepackage{amsmath} % For align, operator 1name
\usepackage{bm}
\usepackage{mathtools}
\usepackage{cases}
\usepackage{nccmath}
\usepackage{mathrsfs}
\usepackage{upgreek}
\usepackage{adjustbox}
\usepackage{tikz}
\usepackage{bigdelim}
\usepackage{comment}
\usepackage{lipsum}
\usepackage{subcaption}
\usepackage{setspace}
\newenvironment{problem*}[1]
{%
    \ifcsname problem@#1\endcsname
    \expandafter\xdef\csname problem@#1\endcsname
    {%
        \the\numexpr\csname problem@#1\endcsname+1\relax
    }%
    \else
    \expandafter\xdef\csname problem@#1\endcsname{1}%
    \fi
    \edef\thesubproblem
    {%
        \getrefnumber{#1}.\csname problem@#1\endcsname
    }%
    \subproblem
}
{
    \endsubproblem
}

\usetikzlibrary{shapes,arrows, intersections, fit}

\usepackage{algorithm}
\usepackage{algpseudocode}

\algdef{SE}[DOWHILE]{Do}{doWhile}{\algorithmicdo}[1]{\algorithmicwhile\ #1}%

\makeatletter
\define@key{Gin}{Trim}
            {\let\Gin@viewport@code\Gin@trim\expandafter\Gread@parse@vp#1 \\}
\makeatother

\newcommand{\R}{\mathbb{R}}
\newcommand{\Z}{\mathbb{Z}}
\newcommand{\N}{\mathbb{N}}
\newcommand{\Sym}{\mathbb{S}}
\newcommand{\E}{\mathbb{E}}
\newcommand{\Ncal}{\mathcal{N}}
\DeclareMathOperator{\tr}{tr}
\DeclareMathOperator{\range}{range}
\DeclareMathOperator{\blkdiag}{blkdiag}
\DeclareMathOperator{\Cov}{Cov}
\newcommand{\independent}{\perp\!\!\!\perp}
\newcommand{\GMM}{\mathrm{GMM}}

\newcommand{\bfe}{\mathbf{e}}
\newcommand{\bfy}{\mathbf{y}}
\newcommand{\bfz}{\mathbf{z}}
\newcommand{\bfx}{\mathbf{x}}

\newcommand{\bfu}{\mathbf{u}}

\newcommand{\bfw}{\mathbf{w}}

\newcommand{\scriptR}{\mathcal{R}}

\newcommand{\scriptG}{\mathcal{G}}

\renewcommand{\Pr}{\mathbb{P}}

\definecolor{c0}{rgb}{0,0.4470,0.7410}       % x_0
\definecolor{cf}{rgb}{0.4660,0.6740,0.1880}  % x_T
\definecolor{cw}{rgb}{0.8500,0.3250,0.0980}  % w_k and J_k

\begin{document}
\title{Gaussian-Mixture Steering of Linear Time-Varying Systems Under\\ Mixture Noise Preview$^{*\dagger}$}

\author{Vignesh Sivaramakrishnan$^1$, Joshua Y. Pilipovsky$^2$, Sean A. Phillips$^3$
\thanks{%
$\hspace{-1.05em}^{*}\hspace{1.1em}$ Approved for public release; distribution is unlimited. Public Affairs approval \#AFRL-2026-4174.\newline
$\dagger\hspace{0.85em}$ 
The views and conclusions contained herein are those of the authors and should not be interpreted as necessarily representing the official policies or endorsements, either expressed or implied, of Air Force Research Laboratory, Space Force, Air Force, and/or the U.S. Government.
$^1\hspace{0.73em}$ National Academies, NRC Postdoctoral Fellow, Space Vehicles Directorate, Air Force Research Laboratory, Kirtland AFB, NM 87117, \texttt{vignesh.sivaramakrishnan.3cc.ext@afresearchlab.com}.
$^2\hspace{1.25em}$ Senior Research Engineer, RTX Technology Research Center (RTRC), East Hartford, CT, 06119, \texttt{joshua.y.pilipovsky@rtx.com}\newline
$^3\hspace{0.5em}$ Technology Advisor, Space Control Branch, Space Warfare Directorate (RJ), Air Force Research Laboratory, Kirtland AFB, NM, 87117, \texttt{sean.phillips.9@spaceforce.mil}.\newline
\indent\indent Vignesh Sivaramakrishnan is supported in part by an appointment to the NRC Research Associateship Program at the Air Force Research Laboratory, Space Vehicles Directorate.
}}
\maketitle

\begin{abstract}
This letter considers distribution steering of a discrete-time linear time-varying system between Gaussian mixture endpoints when the additive disturbance is itself a Gaussian mixture at every timestep.
Under the disturbance preview assumption, the noise realized at each step is known before the input is applied.
The admissible policies are affine, with gain and feedforward indexed by the mixture component of the disturbance.
Synthesis reduces to two convex programs for each pair of boundary mixture components, all coupled by a linear program over the transport weights.
To use our controller, we require the disturbance mixture component to be estimated from realizations of the disturbance.
Therefore, we provide a sub-optimality measure of our approach when disturbance mixture component mismatches occur.
We also give a necessary condition on the terminal covariances attainable under previewed mixture noise.
A 2D double integrator perturbed by a previewed disturbance illustrates the approach.
\end{abstract} 

\section{Introduction}\label{sec:intro}

\begin{figure}[t!]
    \centering
    \includegraphics[width=\linewidth]{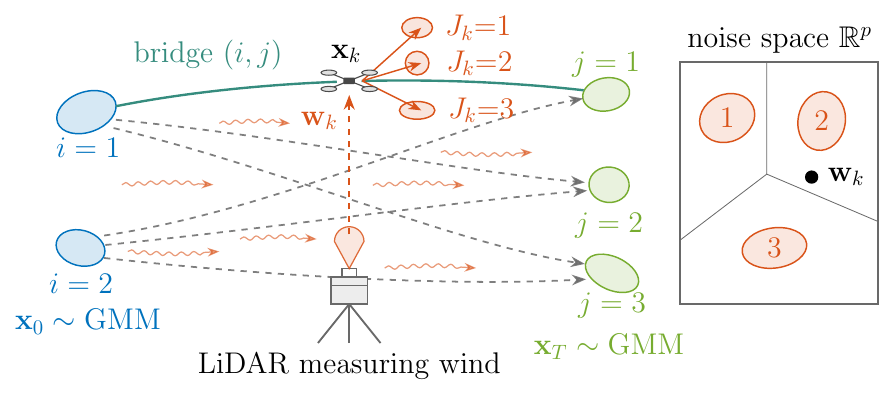}
\vspace{-16pt}
\caption{Steering under previewed mixture noise. A drone fleet moves from \textcolor{c0}{$\mathbf{x}_0\sim\mathrm{GMM}$} to \textcolor{cf}{$\mathbf{x}_T\sim\mathrm{GMM}$} under wind.
At each $k$, a LiDAR previews the disturbance \textcolor{cw}{$\mathbf{w}_k$} before the input is applied, and its label \textcolor{cw}{$J_k$} is estimated from the realization.}
\label{fig:overview}
\end{figure}

This letter addresses the distribution steering problem for discrete-time linear time-varying systems with Gaussian mixture model (GMM) boundary conditions, driven by GMM disturbances whose realization at each step is available before the input commits (Fig.~\ref{fig:overview}). 
This information structure is known as disturbance or noise preview \cite{liu2026stochastic}, and it arises when a sensor measures the disturbance before it acts, as with LiDAR measuring wind on wind turbines \cite{ozdemir2013design,scholbrock2016lidar}, road preview for active suspension \cite{tomizuka1976optimum,hac1992optimal}, and ground-based LiDAR measuring gusts for multirotor hover and landing~\cite{mendez2023wind}. 
Finite GMMs approximate any, potentially multi-modal, density to arbitrary accuracy~\cite{alspach1972nonlinear}, can be fitted to data such as wind measurements~\cite{zhang2022gaussian}, and are a form in which nonlinear filtering, uncertainty propagation, and multi-object tracking deliver their output~\cite{terejanu2008uncertainty,hussein2012aegis,11259237}.

Covariance steering, a form of dynamics-constrained optimal mass transport~\cite{7549018}, seeks a feedback policy that drives a linear stochastic system from an initial Gaussian distribution to a desired terminal one.
The discrete-time problem admits a lossless convex reformulation as a semi-definite program~\cite{bakolas2018finite,liu2024optimal}, with chance constraints addressed in~\cite{okamoto2018optimal}.
Gaussian mixture boundary conditions were introduced in~\cite{balci2024density,kumagai2024chance} without disturbances and in~\cite{rapakoulias2026mixture} with Gaussian, but not GMM, disturbances.
General disturbances have been addressed by characteristic-function~\cite{sivaramakrishnan2022distribution}, power-moment~\cite{wu2024distribution}, and unscented-transformation~\cite{qi2025nongaussian} methods, of which the first accommodates mixture process noise but matches the terminal density approximately, through a non-convex program with a single affine disturbance feedback policy~\cite{goulart2006optimization}.
None combine mixture process noise with a convex synthesis that exploits the mixture structure.
A related line of work steers Markov jump linear systems~\cite{shrivastava2025chance,wang2026mjls}, where the coupled mean and covariance dynamics are handled by an iterative or lifted optimization.

\textit{The main contribution of this letter is a decoupled, convex synthesis that steers a discrete-time linear system between GMM boundary conditions under previewed disturbances modeled as GMMs.}
We index the policy on each mixture component of the disturbance GMM.
When the disturbance component is known, the resulting policy ensures the system's terminal state GMM matches the mean and covariance of every component of the terminal boundary GMM.
We also provide a result on the controller's sub-optimality when disturbance component mismatches occur, since it is estimated from realizations of the disturbance.
Finally, we give a necessary condition on attainable terminal covariances under preview, highlighting the conservatism or over-optimism caused by the lack of its availability.

\section{Preliminaries and Problem Formulation}\label{sec:prelimProbForm}

\subsection{Notation}

The symmetric, positive semi-definite, and positive-definite $n \times n$ matrices are $\Sym^{n}$, $\Sym^{n}_{\succeq 0}$, and $\Sym^{n}_{\succ 0}$, with $\succeq$ the Loewner order.
We set $[M] := \{1, \dots, M\}$, $\Z_{a:b} := \{a, \dots, b\}$, and $s_{a:b} := (s_{a}, \dots, s_{b})$.
Random vectors are bold, $\bfw$, with density $p_{\bfw}$, and component labels are capitals, $J \in [M]$, with realizations $l$.
A Gaussian has law $\Ncal(\mu, \Sigma)$ and pdf $\Ncal(x;\mu, \Sigma)$, and $\bfx \independent \bfy$ denotes independence.
The Frobenius norm is $\|\cdot\|_F$.

\begin{defn}[Gaussian mixture]\label{def:gmm}
Let $M \in \N$, let $\alpha_{\bfw,l} \ge 0$ for $l \in [M]$ with
$\sum_{l \in [M]} \alpha_{\bfw,l} = 1$, and let $\mu_{\bfw,l} \in \R^{p}$,
$\Sigma_{\bfw,l} \in \Sym^{p}_{\succ 0}$. A random vector $\bfw \sim \GMM(\{\alpha_{\bfw,l}, \mu_{\bfw,l},
\Sigma_{\bfw,l}\}_{l \in [M]})$,
if there exists a latent label $J \in [M]$ with
$\Pr(J = l) = \alpha_{\bfw,l}$ such that
$\bfw \mid (J = l) \sim \Ncal(\mu_{\bfw,l}, \Sigma_{\bfw,l})$ for every
$l \in [M]$. This yields the density
$p_{\bfw}(w) = \sum_{l \in [M]} \alpha_{\bfw,l}\,
\Ncal(w; \mu_{\bfw,l}, \Sigma_{\bfw,l})$.
\end{defn}

\vspace{-8pt}
\subsection{Problem Statement}\label{sec:probStatement}
The timestep and horizon are $k\in\N$ and $T\in\N$, respectively.
We restrict ourselves to systems of the form,
\begin{equation}\label{eq:sys}
    \bfx_{k+1} = A_k\bfx_k + B_k \bfu_k + G_k\bfw_k,
\end{equation}
where $A_k\in\R^{n\times n},B_k\in\R^{n\times m},G_k\in\R^{n\times p}$ are
system matrices, $\bfx_k\in\R^{n}$ is the state, and $\bfu_k\in\R^{m}$ is the
input.
The disturbance is $\bfw_{k}\sim\GMM(\{\alpha_{\bfw_k,l}, \mu_{\bfw_k,l}, \Sigma_{\bfw_k,l}\}_{l\in[M_{k}]})$,
with component label $J_{k}\in[M_{k}]$ in the sense of
Definition~\ref{def:gmm}.
\begin{assum}\label{as:noise}
The pairs $\{(J_{k}, \bfw_{k})\}_{k=0}^{T-1}$ are mutually independent and
independent of the initial state $\bfx_{0}$.
\end{assum}
\begin{assum}~\label{as:preview}
At each $k$, the realization of $\bfw_{k}$ is available before the input $\bfu_{k}$ is applied, but not the realization of $J_k$.
\end{assum}
Typically $\bfw_{k}$ comes from the residual $G_{k}\bfw_{k} = \bfx_{k+1} - A_{k}\bfx_{k} - B_{k}\bfu_{k}$, recovered causally \cite{goulart2006optimization} and therefore too late for $\bfu_{k}$ to reject.
Assumption~\ref{as:preview} instead makes $\bfw_{k}$ available before it acts,
as the applications in Section~\ref{sec:intro} highlight.
\begin{assum}~\label{as:unmatched-noise}
    At each $k\in\mathbb{Z}_{0:T-1}$, $\range(G_k) \nsubseteq \range(B_k)$, i.e., no $D_k\in\mathbb{R}^{m\times p}$ satisfies $B_k D_k = G_k$.
\end{assum}
    Assumption~\ref{as:unmatched-noise} rules out full rejection of the previewed disturbance and is common in practice~\cite{yang2012nonlinear,tabassum2021variance}.
    If instead $\range(G_k) \subseteq \range(B_k)$, the input $\bfu_k = \bar\bfu_k - B_k^\dagger G_k \bfw_k$ cancels the noise at $k$ exactly for any nominal input $\bar\bfu_k$.

\begin{assum}~\label{as:ctrb}
    The pairs $\{(A_{k}, B_{k})\}_{k=0}^{T-1}$ are controllable over the horizon, i.e.,\ the controllability Gramian $\scriptG_{T:0} = \sum_{k=0}^{T-1}\Phi_{A}(T, k{+}1) B_{k} B_{k}^{\intercal} \Phi_{A}(T, k{+}1)^{\intercal}$ is non-singular, where $\Phi_{A}(k_{2}, k_{1}) := A_{k_{2}-1}\cdots A_{k_{1}}$ with $\Phi_{A}(k, k) := I_{n}$~\cite[Assumption 1]{balci2024density}.
\end{assum}
\begin{assum}~\label{as:invertible}
    At each $k\in\mathbb{Z}_{0:T-1}$, $A_k$ is invertible~\cite[Assumption 1]{liu2024optimal}.
\end{assum}

We restrict attention to affine control policies of the form,
\begin{equation}\label{eq:inputLinearGeneric}
  \pi_{k}(\bfx_{k}, J_k) = K_{k}(J_k)\left(\bfx_{k} -
  \mu_{\bfx_k}\right) + v_{k}(J_k),
\end{equation}
with gain $K_k(J_k)\in\mathbb{R}^{m\times n}$, state mean
$\mu_{\bfx_k}\in\mathbb{R}^n$, and feedforward control $v_k(J_k)\in\mathbb{R}^m$,
indexed by $J_{k}$.
We minimize the following quadratic cost,
\begin{equation}\label{eq:cost}
    C(\pi) = \sum_{k=0}^{T-1}
      \E\!\left[\bfe_{k}^{\intercal}Q_{k}\bfe_{k} +
      \bfu_{k}^{\intercal}R_{k}\bfu_{k}\right],
\end{equation}
where $\bfe_{k} := \bfx_{k} - x_{d,k}$ is the deviation of the state from a
given reference sequence $\{x_{d,k}\}_{k=0}^{T-1}\subset\R^{n}$, and
$Q_{k}\in\Sym^{n}_{\succ 0}$, $R_{k}\in\Sym^{m}_{\succ 0}$ weight the deviation and the
input, respectively.
A policy $\pi := \{\pi_{k}\}_{k=0}^{T-1}$ is the sequence of affine maps in
\eqref{eq:inputLinearGeneric}.
We present the optimization problem we wish to solve.
\begin{problem}\label{prob:main}
With Assumptions~\ref{as:noise}--\ref{as:invertible}, find a policy $\pi$ that assigns the state to a pair of an initial component $i\in[N_0]$ and a terminal component $j\in[N_f]$, applies to it a policy $\pi^{ij}$ of the form \eqref{eq:inputLinearGeneric}, and solves,

\begin{subequations}\label{eq:prob}
\begin{align}
  \mathrm{minimize} \quad
    & C(\pi) \label{eq:prob-obj}\\
  \mathrm{subject\ to} \quad
    & \eqref{eq:sys},\ \bfx_{0}\sim\GMM\big(\{\alpha^{0}_{i},\mu^{0}_{i},\Sigma^{0}_{i}\}_{i\in[N_{0}]}\big), \label{eq:prob-dyn}\\
    & \bfx_{T}\sim\textstyle\sum_{j\in[N_{f}]}\alpha^{f}_{j}\,p_{j}, \label{eq:prob-tl}\\
    & \E[\bfx_{T}\,|\,i,j]=\mu^{f}_{j},\ \Cov(\bfx_{T}\,|\,i,j)=\Sigma^{f}_{j}. \label{eq:prob-tc}
\end{align}
\end{subequations}
\end{problem}
Constraint \eqref{eq:prob-tc} fixes the mean and covariance of each terminal component $p_j$, and not its shape.
Matching these moments for every pair is sufficient, but not necessary, for matching them for every $p_j$.
Conditioned on the pair and on the noise labels $J_{0:T-1}$, the closed loop is affine and driven by Gaussian variables, so $p_j$ is a GMM over the $\prod_{k\in\Z_{0:T-1}}M_k$ label histories, in contrast to \cite{rapakoulias2026mixture}, where Gaussian disturbances keep each component of the terminal state GMM a single Gaussian.
Problem~\ref{prob:main} reduces to the single Gaussian setting of \cite{rapakoulias2026mixture} under Gaussian disturbances when $M_k = 1$, and Section~\ref{sec:law} states when each component collapses to a single Gaussian under mixture noise.

\section{Control Law Accounting for\\ Previewed Gaussian Mixture Process Noise}\label{sec:method}
It is well known that the control law $\bfu_{k} = \pi_{k}(\bfx_{k}) = K_{k}\left(\bfx_{k} - \mu_{\bfx_k}\right) + v_{k}$, akin to \eqref{eq:inputLinearGeneric}, steers a linear time-varying system from a Gaussian to another Gaussian under Gaussian disturbances~\cite{liu2024optimal}.
We compute the feedforward and gain terms via a quadratic program (QP) and a semi-definite program (SDP), respectively.
The following  steers the mean,

\begin{subequations}\label{eq:QPmean}
\begin{align}
  \underset{v_k}{\mathrm{min.}}
    &\ \Tilde{C}_{\mu}(\pi) = \sum_{k=0}^{T-1}\mu_{\bfe_k}^{\intercal}Q_k\mu_{\bfe_k} +  v_k^{\intercal}R_k v_k,\\
  \mathrm{s.t.} &\ \mu_{\bfx_{k+1}} \hspace{-0.3em}= A_k\mu_{\bfx_{k}} + B_kv_k + G_k \mu_{\bfw_k}, k\in\Z_{0:T-1},\label{eq:meanState}\\
  &\ \mu_{\bfx_0} = \mu^0, \mu_{\bfx_T} = \mu^f.
\end{align}
\end{subequations}
The second optimization problem steers the covariance,

\begin{subequations}\label{eq:SDPcov}
\begin{align}
  \underset{U_k,Y_k}{\mathrm{min.}}
    &\hspace{0.2in} \Tilde{C}_{\Sigma}(\pi) = \sum_{k=0}^{T-1}\left[\tr(Q_k\Sigma_{\bfx_k})+ \tr(R_kY_k)\right],\\
  \mathrm{s.t.} & \hspace{0.154in} \Sigma_{\bfx_{k+1}} = A_k\Sigma_{\bfx_k}A_k^{\intercal} + B_kU_kA_k^{\intercal} + A_kU_k^{\intercal}B_k^{\intercal} + \nonumber\\
  &\hspace{1.39in} B_kY_kB_k^{\intercal} + G_k\Sigma_{\bfw_k}G_k^{\intercal},\label{eq:covState}\\
  &\quad 
  \begin{bmatrix}
      \Sigma_{\bfx_k} & U^{\intercal}_k\\
      U_k & Y_k
  \end{bmatrix}\succeq 0, k\in\Z_{0:T-1},\label{eq:lmiK}\\
  &\quad \Sigma_{\bfx_{0}} = \Sigma^0, \Sigma_{\bfx_T} = \Sigma^f. 
\end{align}
\end{subequations}%
Note $\mu_{\bfe_k} = \mu_{\bfx_k} - x_{d,k}$, $\bfx_0\sim\mathcal{N}(\mu^0,\Sigma^0)$, and $\bfx_T\sim\mathcal{N}(\mu^f,\Sigma^f)$.
We recover the gain via $K_k = U_k\Sigma_{\bfx_k}^{-1}$.

GMM steering applies \eqref{eq:QPmean} and \eqref{eq:SDPcov} between one component of the initial GMM and one of the terminal GMM~\cite{balci2024density,rapakoulias2026mixture}.
We solve a linear program (LP),
\begin{subequations}\label{eq:gmmLP}
\begin{align}
  \underset{\lambda_{ij}\geq 0}{\mathrm{min.}}
    &\quad \textstyle\sum_{i\in[N_0]}\sum_{j\in[N_f]} \lambda_{ij}\Tilde{C}^{ij}(\pi),\label{eq:gmmLP-obj}\\
  \mathrm{s.t.} &\quad \textstyle\sum_{j\in[N_f]}\lambda_{ij} = \alpha_i^0,\ \sum_{i\in[N_0]}\lambda_{ij} = \alpha_j^f,\label{eq:gmmLP-con}
\end{align}
\end{subequations}
where $\lambda_{ij}$ denotes the fraction of mass assigned from initial, $i$, to terminal, $j$, mixture component, and $\Tilde{C}^{ij}(\pi) = \Tilde{C}^{ij}_{\mu}(\pi) + \Tilde{C}^{ij}_{\Sigma}(\pi)$ is the sum of optimal costs from \eqref{eq:QPmean} and \eqref{eq:SDPcov} with boundary distributions $\Ncal(\mu^0_i,\Sigma^0_i)$ and $\Ncal(\mu^f_j,\Sigma^f_j)$.
The constraints \eqref{eq:gmmLP-con} enforce preservation of mass between initial and terminal GMMs.
Since we solve  \eqref{eq:QPmean} and \eqref{eq:SDPcov} independently of the assignment, $\Tilde{C}^{ij}(\pi)$ is fixed for \eqref{eq:gmmLP}.
Thus, minimizing \eqref{eq:gmmLP-obj} places mass on the cheapest pairs.

The objective \eqref{eq:gmmLP-obj} equals the cost \eqref{eq:cost} when the assigned pair is random with distribution $\lambda_{ij}$. Conditioning the expectation in \eqref{eq:cost} on the pair gives $\Tilde{C}^{ij}(\pi)$, and the tower property yields $C(\pi) = \sum_{ij}\lambda_{ij}\Tilde{C}^{ij}(\pi)$.
Two constructions in the literature achieve this with the controller,
\begin{equation}\label{eq:pairPolicy}
    \bfu^{ij}_k = K^{ij}_k(\bfx_k - \mu^{ij}_{\bfx_k}) + v^{ij}_k,
    \qquad (i,j)\sim\gamma_{ij}(\cdot),
\end{equation}
differing only in the probability law $\gamma_{ij}(\cdot)$.
In \cite{balci2024density}, the pair is drawn once at $k=0$, with
\begin{equation}\label{eq:initRule}
\gamma_{ij}(x_0) = \frac{\lambda_{ij}\Ncal(x_0;\mu^0_i,\Sigma^0_i)}{\sum_{i'=1}^{N_0}\alpha^0_{i'}\Ncal(x_0;\mu^0_{i'},\Sigma^0_{i'})}.
\end{equation}
In \cite{rapakoulias2026mixture}, the pair is redrawn at every step, with
\begin{equation}\label{eq:stepRule}
\gamma_{ij}(x_k) =\frac{ \lambda_{ij}p^{ij}_k(x_k)}{\sum_{i'=1}^{N_0}\sum_{j'=1}^{N_f}\lambda_{i'j'}p^{i'j'}_k(x_k)},
\end{equation}
where $p^{ij}_k$ is the state pdf at time $k$ under the pair $(i,j)$, and \eqref{eq:prob-tc} then refers to the pair drawn at $k=T-1$.
When the disturbance is a previewed GMM, we modify \eqref{eq:pairPolicy} to accept its component labels.
Section~\ref{sec:law} indexes the gain and feedforward by the label $J_k$.
Thus Section~\ref{sec:runtime} estimates $J_k$ from realizations of $\bfw_k$.
    
\subsection{Control Law Synthesis}\label{sec:law}
We solve \eqref{eq:prob} using the control law,
\begin{equation}\label{eq:mixtureControl}
    \bfu^{ij}_k = K^{ij}_k(J_k)\big(\bfx_k - \mu^{ij}_{\bfx_k}\big) + v^{ij}_k + \Tilde{v}^{ij}_k(J_k),
    \ (i,j)\sim\gamma_{ij}(\cdot),
\end{equation}
in which the gain $K^{ij}_k$ and the \emph{deviation feedforward} $\Tilde{v}^{ij}_k$ are indexed by the component label $J_k\in[M_k]$ of the noise $\bfw_k$.
The feedforward is split into a label-independent part $v^{ij}_k$ and a deviation $\Tilde{v}^{ij}_k(J_k)$ normalized so that
\begin{equation}\label{eq:zeroMean}
    \sum_{l\in[M_k]}\alpha_{\bfw_k,l}\Tilde{v}^{ij}_k(l) = 0_m.
\end{equation}
Normalization \eqref{eq:zeroMean} incurs no loss of generality. For any label-dependent feedforward, $v^{ij}_k$ is its mixture average and $\Tilde{v}^{ij}_k(l)$ the residual.
Substituting \eqref{eq:mixtureControl} into \eqref{eq:sys} yields
\begin{align}\label{eq:closedLoop}
    \bfx_{k+1} \hspace{-0.1em}= \bar{A}^{ij}_k(J_k)\bfx_k
      - B_kK^{ij}_k(J_k)\mu^{ij}_{\bfx_k}+ \Delta v + G_k\bfw_k,
\end{align}
where $\bar{A}^{ij}_k(l) := A_k + B_kK^{ij}_k(l)$ is the closed-loop matrix under component $l$ and $\Delta v = B_k\big(v^{ij}_k + \Tilde{v}^{ij}_k(J_k)\big)$.

The mean recursion follows from the expectation of \eqref{eq:closedLoop},
\begin{align}\label{eq:meanExp}
    \mu^{ij}_{\bfx_{k+1}} &= A_k\mu^{ij}_{\bfx_k}
      + B_k\Delta K\nonumber\\
      &\quad + B_k\big(v^{ij}_k + \E[\Tilde{v}^{ij}_k(J_k)]\big) + G_k\E[\bfw_k],
\end{align}
where $\Delta K = \E\big[K^{ij}_k(J_k)\big(\bfx_k - \mu^{ij}_{\bfx_k}\big)\big]$ in which both label-dependent terms vanish.
The state $\bfx_k$ is a function of $\bfx_0$ and $\bfw_{0:k-1}$, while $J_k$ labels $\bfw_k$, so Assumption~\ref{as:noise} gives $J_k \independent \bfx_k$.
Applying the tower property, $\E[\,\cdot\,] = \E\big[\E[\,\cdot \mid J_k]\big]$, the gain equals the constant $K^{ij}_k(l)$ on $\{J_k = l\}$ and factors out of the inner expectation, leaving
\begin{align}\label{eq:fbTower}
    \Delta K =  \sum_{l\in[M_k]}\alpha_{\bfw_k,l}K^{ij}_k(l)
      \big(\E[\bfx_k \mid J_k = l] - \mu^{ij}_{\bfx_k}\big).
\end{align}
Independence, i.e., $J_k \independent \bfx_k$, gives $\E[\bfx_k \mid J_k = l] = \mu^{ij}_{\bfx_k}$ for every $l\in[M_k]$, so each summand of \eqref{eq:fbTower} is zero and the feedback contributes nothing to the mean.
The deviation feedforward vanishes by \eqref{eq:zeroMean}, which is what makes $v^{ij}_k$ the mean feedforward.
With $\E[\bfw_k] = \bar\mu_{\bfw_k} := \sum_{l\in[M_k]}\alpha_{\bfw_k,l}\mu_{\bfw_k,l}$, \eqref{eq:meanExp} reduces to
\begin{equation}\label{eq:meanRec}
    \mu^{ij}_{\bfx_{k+1}} = A_k\mu^{ij}_{\bfx_k} + B_kv^{ij}_k + G_k\bar\mu_{\bfw_k},
\end{equation}
which is \eqref{eq:meanState} with $\mu_{\bfw_k} = \bar\mu_{\bfw_k}$.
We can therefore solve the mean steering problem with the same QP, \eqref{eq:QPmean}.
The normalization \eqref{eq:zeroMean} keeps the mean and covariance problems separate, and is enforced in the covariance steering problem.

The covariance recursion follows from $\Sigma^{ij}_{\bfx_{k+1}} = \E[(\bfx_{k+1} - \mu^{ij}_{\bfx_{k+1}})(\bfx_{k+1} - \mu^{ij}_{\bfx_{k+1}})^{\intercal}]$, so we require the dynamics of the deviation $\bfx_{k+1} - \mu^{ij}_{\bfx_{k+1}}$.
Subtracting \eqref{eq:meanRec} from \eqref{eq:closedLoop}, the feedforward $B_kv^{ij}_k$ cancels and the terms in $\bfx_k$ and $\mu^{ij}_{\bfx_k}$ collect, leaving
\begin{align}\label{eq:devRaw}
    &\bfx_{k+1} - \mu^{ij}_{\bfx_{k+1}}
      = \bar{A}^{ij}_k(J_k)\big(\bfx_k - \mu^{ij}_{\bfx_k}\big) + B_k\Tilde{v}^{ij}_k(J_k) + \Delta \bfw_k.
\end{align}
The term $\Delta \bfw_k = G_k\bfw_k - G_k\bar\mu_{\bfw_k}$ is neither zero mean nor deterministic conditioned on $J_k$, as component $J_k$ has mean $\mu_{\bfw_k,J_k}$, not $\bar\mu_{\bfw_k}$.
Introducing $G_k\mu_{\bfw_k,J_k}$ separates them,
\begin{equation}\label{eq:noiseSplit}
    \Delta \bfw_k
      = \underbrace{G_k\big(\mu_{\bfw_k,J_k} - \bar\mu_{\bfw_k}\big)}_{\text{deterministic given }J_k}
      + \underbrace{G_k\big(\bfw_k - \mu_{\bfw_k,J_k}\big)}_{\text{zero mean given }J_k},
\end{equation}
and collecting the deterministic part with $B_k\Tilde{v}^{ij}_k(J_k)$ into,
\begin{subequations}\label{eq:delta}
\begin{align}
    d_k(l) &:= G_k\big(\mu_{\bfw_k,l} - \bar\mu_{\bfw_k}\big), \label{eq:offset}\\
    \delta^{ij}_k(l) &:= B_k\Tilde{v}^{ij}_k(l) + d_k(l), \label{eq:spread}
\end{align}
\end{subequations}
casts \eqref{eq:devRaw} as,
\begin{align}\label{eq:devRec}
    \bfx_{k+1} - \mu^{ij}_{\bfx_{k+1}} = \bar{A}^{ij}_k(J_k)\big(\bfx_k - \mu^{ij}_{\bfx_k}\big)
      + \delta^{ij}_k(J_k) +\nonumber\\
      G_k\big(\bfw_k - \mu_{\bfw_k,J_k}\big).
\end{align}
Since $J_k \independent \bfx_k$ and $\E[\bfw_k \mid J_k = l] = \mu_{\bfw_k,l}$, the first and third terms of \eqref{eq:devRec} are conditionally zero mean, so
\begin{equation}\label{eq:condMean}
    \E\big[\bfx_{k+1} \mid J_k = l\big] - \mu^{ij}_{\bfx_{k+1}} = \delta^{ij}_k(l).
\end{equation}
The quantity \eqref{eq:spread} is thus the displacement of the label-conditioned mean from the unconditional one, which we call the \emph{between-label mean spread}.
Subtracting \eqref{eq:condMean} from \eqref{eq:devRec} removes the deterministic term, leaving $\bfz^{ij}_k(l) := \bar{A}^{ij}_k(l)(\bfx_k - \mu^{ij}_{\bfx_k}) + G_k(\bfw_k - \mu_{\bfw_k,l})$, so that $\Cov(\bfx_{k+1} \mid J_k = l) = \E[\bfz^{ij}_k(l)\bfz^{ij}_k(l)^{\intercal} \mid J_k = l]$.
Expanding this product, the cross terms vanish by Assumption~\ref{as:noise}, while
$\E[(\bfx_k - \mu^{ij}_{\bfx_k})(\bfx_k - \mu^{ij}_{\bfx_k})^{\intercal} \mid J_k = l] = \Sigma^{ij}_{\bfx_k}$
by $J_k \independent \bfx_k$ and
$\E[(\bfw_k - \mu_{\bfw_k,l})(\bfw_k - \mu_{\bfw_k,l})^{\intercal} \mid J_k = l] = \Sigma_{\bfw_k,l}$
by Definition~\ref{def:gmm}, leaving
\begin{equation}\label{eq:covCond}
    \Cov\big(\bfx_{k+1} \mid J_k = l\big)
      = \bar{A}^{ij}_k(l)\Sigma^{ij}_{\bfx_k}\bar{A}^{ij}_k(l)^{\intercal}
      + G_k\Sigma_{\bfw_k,l}G_k^{\intercal}.
\end{equation}
A single $\Sigma^{ij}_{\bfx_k}$ serves every label in \eqref{eq:covCond} because the state at time $k$ does not depend on the current mixture component.
Equations \eqref{eq:condMean} and \eqref{eq:covCond} give the structure of $p_j$ deferred from Section~\ref{sec:probStatement}.
On a fixed label history, \eqref{eq:devRec} is affine, so $\bfx_T$ is Gaussian, and averaging over the $\prod_{k\in\Z_{0:T-1}}M_k$ histories leaves $p_j$ a GMM.
Its components coincide, and $p_j$ is a single Gaussian as in \cite{rapakoulias2026mixture}, when \eqref{eq:covCond} is label-independent and $\Tilde{v}^{ij}_k(l)$ cancels \eqref{eq:spread}, which requires $d_k(l)\in\range(B_k)$.

The law of total variance gives $\Sigma^{ij}_{\bfx_{k+1}} = \E[\Cov(\bfx_{k+1} \mid J_k)] + \Cov(\E[\bfx_{k+1} \mid J_k])$, with all moments conditional on the fixed pair $(i,j)$.
The first term is the $\alpha_{\bfw_k,l}$ weighted average of \eqref{eq:covCond}.
For the second, \eqref{eq:condMean} places the conditional means at $\mu^{ij}_{\bfx_{k+1}} + \delta^{ij}_k(l)$, which average to $\mu^{ij}_{\bfx_{k+1}}$ by \eqref{eq:zeroMean} and $\bar\mu_{\bfw_k}$, so it is $\textstyle\sum_{l\in[M_k]}\alpha_{\bfw_k,l}\delta^{ij}_k(l)\delta^{ij}_k(l)^{\intercal}$.
With \eqref{eq:covCond},
\begin{align}\label{eq:covRec}
    &\Sigma^{ij}_{\bfx_{k+1}} \hspace{-0.5em}= \hspace{-0.5em}\sum_{l\in[M_k]}\hspace{-0.5em}\alpha_{\bfw_k,l}\Big[
      \bar{A}^{ij}_k(l)\Sigma^{ij}_{\bfx_k}\bar{A}^{ij}_k(l)^{\intercal}
      + G_k\Sigma_{\bfw_k,l}G_k^{\intercal}
      + \nonumber\\
      &\hspace{16em}\delta^{ij}_k(l)\delta^{ij}_k(l)^{\intercal}\Big].
\end{align}%
Within the brackets, the first two terms reproduce the recursion in \eqref{eq:covState} for component $l$, and the third is the between-label mean spread, which arises when the noise is a mixture and is the first key departure from~\cite{balci2024density,rapakoulias2026mixture}.

The recursion \eqref{eq:covRec} is not convex.
In \eqref{eq:SDPcov}, the change of variables $U_k = K_k\Sigma_{\bfx_k}$ with $Y_k \succeq U_k\Sigma_{\bfx_k}^{-1}U_k^{\intercal}$ resolves this through \eqref{eq:lmiK}~\cite[Thm.~5]{liu2024optimal}.
Applied per component in \eqref{eq:covRec}, it leaves the between-label mean spread, which is quadratic in $\Tilde{v}^{ij}_k(l)$ through \eqref{eq:spread}.
Relaxing that separately would go unpenalized by \eqref{eq:cost} and could absorb positive semi-definite slack in the recursion, relaxing $\Sigma^{ij}_{\bfx_T} = \Sigma^f_j$ to an inequality.
We instead absorb the spread into the closed-loop matrix.
Let $\zeta^{ij}_k := [(\bfx_k - \mu^{ij}_{\bfx_k})^{\intercal}\ \ 1]^{\intercal}$ and
\begin{equation}\label{eq:hatData}
  \hat{A}_k(l) := \begin{bmatrix} A_k & d_k(l)\\ 0 & 1\end{bmatrix},\
  \hat{B}_k := \begin{bmatrix} B_k\\ 0\end{bmatrix},\
  \hat{G}_k := \begin{bmatrix} G_k\\ 0\end{bmatrix},
\end{equation}%
so that \eqref{eq:devRec} becomes $\zeta^{ij}_{k+1} = \big(\hat{A}_k(l) + \hat{B}_k\hat{K}^{ij}_k(l)\big)\zeta^{ij}_k + \hat{G}_k(\bfw_k - \mu_{\bfw_k,l})$ on $\{J_k = l\}$, with $\hat{K}^{ij}_k(l) := [\,K^{ij}_k(l)\ \ \Tilde{v}^{ij}_k(l)\,]$.
Since $\bfx_k - \mu^{ij}_{\bfx_k}$ is zero mean, $\hat\Sigma^{ij}_k := \E[\zeta^{ij}_k\zeta^{ij\intercal}_k] = \blkdiag(\Sigma^{ij}_{\bfx_k},1)$, and $\hat{U}^{ij}_k(l) := \hat{K}^{ij}_k(l)\hat\Sigma^{ij}_k = [\,U^{ij}_k(l)\ \ \Tilde{v}^{ij}_k(l)\,]$ with $\hat{Y}^{ij}_k(l) \succeq \hat{U}^{ij}_k(l)(\hat\Sigma^{ij}_k)^{-1}\hat{U}^{ij}_k(l)^{\intercal}$ covers both the gain and the spread.
This lifting is the second key departure from both GMM steering approaches in~\cite{balci2024density,rapakoulias2026mixture}.

The cost \eqref{eq:cost} separates into mean and covariance components, with cross terms vanishing due to \eqref{eq:zeroMean} and the underlying controller \eqref{eq:mixtureControl}, yielding $\Tilde{C}^{ij}(\pi) = \Tilde{C}^{ij}_{\mu}(\pi) + \Tilde{C}^{ij}_{\text{mix}}(\pi)$.
The covariance component of the cost reduces to $\tr(R_k\hat{Y}^{ij}_k(l))$, since $\hat{U}^{ij}_k(l)(\hat\Sigma^{ij}_k)^{-1}\hat{U}^{ij}_k(l)^{\intercal} = U^{ij}_k(l)(\Sigma^{ij}_{\bfx_k})^{-1}U^{ij}_k(l)^{\intercal} + \Tilde{v}^{ij}_k(l)\Tilde{v}^{ij}_k(l)^{\intercal}$.
Thus \eqref{eq:covState} is replaced by \eqref{eq:covRec} in lifted form,

\begin{subequations}\label{eq:SDPcovMixture}
\begin{align}
  \mathrm{min.} &\quad \Tilde{C}^{ij}_{\text{mix}}(\pi) = \sum_{k=0}^{T-1}\Big[\tr(Q_k\Sigma^{ij}_{\bfx_k}) + \bar{C}^{ij}_{\Sigma,k}\Big],\\
  \mathrm{s.t.} &\quad \blkdiag\big(\Sigma^{ij}_{\bfx_{k+1}},1\big)
     = \sum_{l\in[M_k]}\alpha_{\bfw_k,l}\Big[\hat{A}_k(l)\hat\Sigma^{ij}_k\hat{A}_k(l)^{\intercal} \nonumber\\
  &\quad + \hat{B}_k\hat{U}^{ij}_k(l)\hat{A}_k(l)^{\intercal}
     + \hat{A}_k(l)\hat{U}^{ij}_k(l)^{\intercal}\hat{B}_k^{\intercal} \nonumber\\
  &\quad + \hat{B}_k\hat{Y}^{ij}_k(l)\hat{B}_k^{\intercal}
     + \hat{G}_k\Sigma_{\bfw_k,l}\hat{G}_k^{\intercal}\Big],\label{eq:covRecLifted}\\
  &\ \begin{bmatrix}
      \hat\Sigma^{ij}_k & \hat{U}^{ij}_k(l)^{\intercal}\\
      \hat{U}^{ij}_k(l) & \hat{Y}^{ij}_k(l)
  \end{bmatrix}\hspace{-0.2em}\succeq 0, k\in\Z_{0:T-1}, l\in[M_k],\label{eq:lmiHat}\\
  &\quad \eqref{eq:zeroMean},\ \Sigma^{ij}_{\bfx_{0}} = \Sigma^0_i,\ \Sigma^{ij}_{\bfx_T} = \Sigma^f_j,
\end{align}
\end{subequations}
where $\Sigma^{ij}_{\bfx_k}$, $\hat{U}^{ij}_k(l)$, $\hat{Y}^{ij}_k(l)$ are its decision variables and $\bar{C}^{ij}_{\Sigma,k} = \sum_{l\in[M_k]}\alpha_{\bfw_k,l}\tr(R_k\hat{Y}^{ij}_k(l))$.
We list \eqref{eq:zeroMean} in \eqref{eq:SDPcovMixture} even though the $(1,2)$ block of \eqref{eq:covRecLifted} vanishes under it, since that block implies \eqref{eq:zeroMean} only when $B_k$ has full column rank.
Given a solution, $K^{ij}_k(l) = U^{ij}_k(l)(\Sigma^{ij}_{\bfx_k})^{-1}$.
The inequality \eqref{eq:lmiHat} relaxes the equality $\hat{Y}^{ij}_k(l) = \hat{U}^{ij}_k(l)(\hat\Sigma^{ij}_k)^{-1}\hat{U}^{ij}_k(l)^{\intercal}$ and is the only relaxation in the synthesis.
We now show that it is tight and that the solutions of \eqref{eq:QPmean}, \eqref{eq:SDPcovMixture}, and \eqref{eq:gmmLP} assemble into a policy that is feasible for Problem~\ref{prob:main}.
\begin{prop}\label{prop:feasible}
    Let Assumption~\ref{as:invertible} hold and $\alpha_{\bfw_k,l} > 0$, and let \eqref{eq:QPmean} be feasible and \eqref{eq:SDPcovMixture} strictly feasible for every pair $(i,j)$.
    Then \eqref{eq:lmiHat} is tight at every optimum of \eqref{eq:SDPcovMixture} with $\Sigma^{ij}_{\bfx_k}\succ 0$, \eqref{eq:gmmLP} has an optimum $\lambda^{\star}_{ij}$, and the policy \eqref{eq:mixtureControl} attains \eqref{eq:prob-tc}.
\end{prop}
\begin{proof}
    We first fix the pair $(i,j)$.
    Let $\Lambda_{k+1}$ be the multiplier of the upper-left block of \eqref{eq:covRecLifted} and $\Omega_k(l)\succeq 0$ the lower-right block of the multiplier of \eqref{eq:lmiHat}, which exist by strict feasibility.
    Stationarity in $\hat{Y}_k(l)$ and $\hat{U}_k(l)$, with complementary slackness, gives $\Omega_k(l) = \alpha_{\bfw_k,l}(R_k + B_k^{\intercal}\Lambda_{k+1}B_k)$ and $\alpha_{\bfw_k,l}B_k^{\intercal}\Lambda_{k+1}A_k = -\Omega_k(l)U_k(l)\Sigma_{\bfx_k}^{-1}$, so $\alpha_{\bfw_k,l}R_k = \Omega_k(l)(I + U_k(l)\Sigma_{\bfx_k}^{-1}A_k^{-1}B_k)$.
    Since $R_k\succ 0$, $\Omega_k(l)$ is nonsingular, and complementary slackness, $\Omega_k(l)(\hat{Y}_k(l) - \hat{U}_k(l)\hat\Sigma_k^{-1}\hat{U}_k(l)^{\intercal}) = 0$, gives equality, as in~\cite[Thm.~5]{liu2024optimal}.
    Tightness makes the recursion of \eqref{eq:SDPcovMixture} coincide with \eqref{eq:covRec} at the recovered gains and feedforwards, so induction from $\Sigma_{\bfx_0} = \Sigma^0_i$ gives $\Cov(\bfx_T\mid i,j) = \Sigma^f_j$, and \eqref{eq:meanRec} gives $\E[\bfx_T\mid i,j] = \mu^f_j$.
    Feasibility makes each $\Tilde{C}^{ij}(\pi)$ finite, so $\lambda^{\star}_{ij}$ exists and its column sums are the weights $\alpha^f_j$.
\end{proof}
Solving \eqref{eq:QPmean} and \eqref{eq:SDPcovMixture} for each pair and then \eqref{eq:gmmLP} is therefore optimal for Problem~\ref{prob:main} under \eqref{eq:initRule} or \eqref{eq:stepRule}, since \eqref{eq:prob-tc} constrains each pair separately and $C(\pi) = \sum_{ij}\lambda_{ij}\Tilde{C}^{ij}(\pi)$.
However, feasibility of \eqref{eq:SDPcovMixture} is not automatic and Section~\ref{sec:reach} screens for reachable $\Sigma^f_j$.
In addition, sub-optimality will arise when the disturbance mixture components are mis-labeled.

\subsection{Estimating $J_k$, Runtime, and Mismatch Sub-optimality}\label{sec:runtime}
\vspace{-13pt}

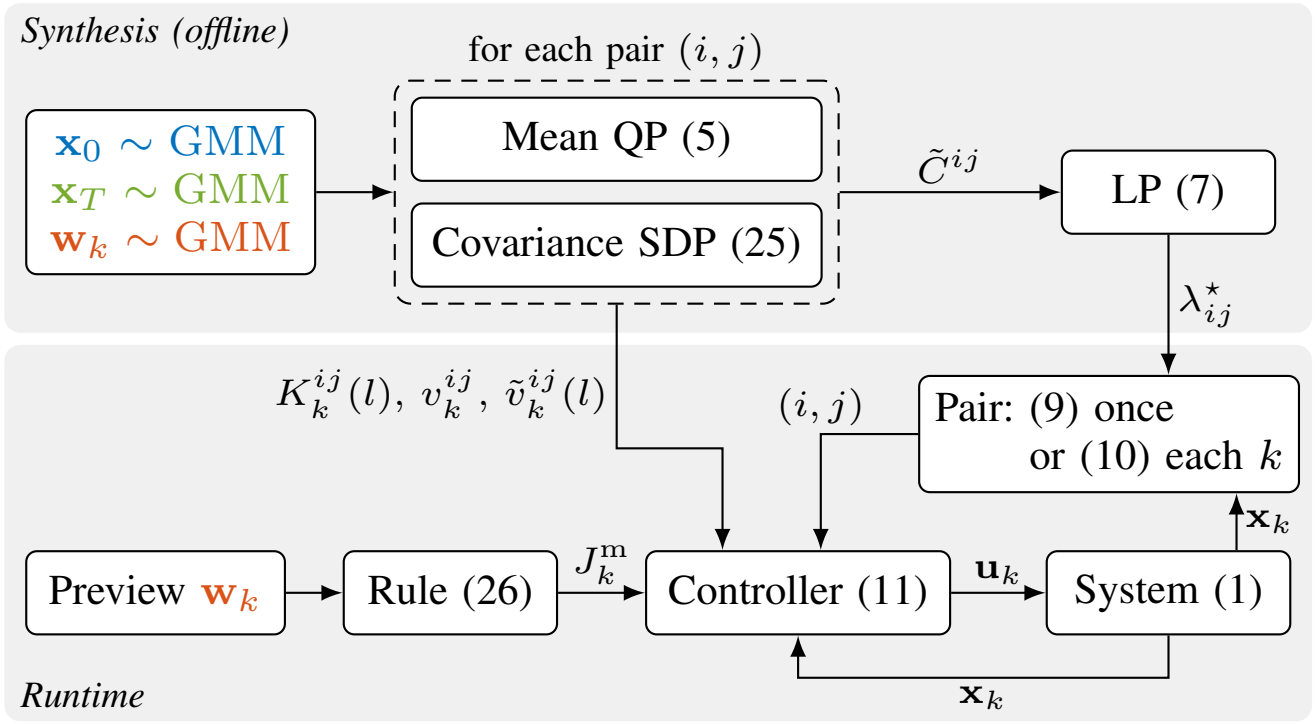
\begin{figure}[ht!]
\centering
\begin{tikzpicture}[
    font=\footnotesize, >=latex, line width=0.4pt,
    blk/.style={draw, fill=white, rounded corners=2pt, align=center, inner sep=3pt, minimum height=5.5mm},
    lbl/.style={font=\scriptsize, inner sep=1.5pt},
    band/.style={font=\scriptsize\itshape, inner sep=1.5pt}]
  % offline and runtime bands
  \fill[black!6, rounded corners=3pt] (-0.15,2.27) rectangle (8.5,4.45);
  \fill[black!6, rounded corners=3pt] (-0.15,-0.3) rectangle (8.5,2.19);
  \node[band, anchor=north west] at (-0.1,4.42) {Synthesis (offline)};
  \node[band, anchor=south west] at (-0.1,-0.27) {Runtime};
  % synthesis
  \node[blk, text width=17mm] (data) at (0.95,3.2)
      {\textcolor{c0}{$\bfx_0\sim\mathrm{GMM}$}\\ \textcolor{cf}{$\bfx_T\sim\mathrm{GMM}$}\\ \textcolor{cw}{$\bfw_k\sim\mathrm{GMM}$}};
  \node[blk, text width=25mm] (qp)  at (3.9,3.55) {Mean QP \eqref{eq:QPmean}};
  \node[blk, text width=25mm] (sdp) at (3.9,2.85) {Covariance SDP \eqref{eq:SDPcovMixture}};
  \node[draw, densely dashed, rounded corners=3pt, fit=(qp)(sdp), inner sep=3pt,
        label={[lbl]above:for each pair $(i,j)$}] (pair) {};
  \node[blk, text width=12mm] (lp) at (7.55,3.2) {LP \eqref{eq:gmmLP}};
  \draw[->] (data) -- (pair);
  \draw[->] (pair) -- node[lbl, above] {$\Tilde{C}^{ij}$} (lp);
  % runtime
  \node[blk, text width=15mm] (prev) at (0.85,0.55) {Preview \textcolor{cw}{$\bfw_k$}};
  \node[blk, text width=12mm] (rule) at (2.8,0.55) {Rule \eqref{eq:argmax}};
  \node[blk, text width=18mm] (law)  at (5.1,0.55) {Controller \eqref{eq:mixtureControl}};
  \node[blk, text width=14mm] (sys)  at (7.55,0.55) {System \eqref{eq:sys}};
  \node[blk, align=left, anchor=east] (draw) at (8.42,1.6) {Pair: \eqref{eq:initRule} once\\ \phantom{Pair: }or \eqref{eq:stepRule} each $k$};
  \draw[->] (prev) -- (rule);
  \draw[->] (rule) -- node[lbl, above] {$J^{\mathrm{m}}_k$} (law);
  \draw[->] (law) -- node[lbl, above] {$\bfu_k$} (sys);
  \draw[->] (sys.south) -- ++(0,-0.28) -| node[lbl, pos=0.25, below] {$\bfx_k$} (law.south);
  \draw[->] ([xshift=4.5mm]sys.north) -- node[lbl, right] {$\bfx_k$} ([xshift=4.5mm]sys.north |- draw.south);
  \draw[->] (lp) -- node[lbl, right] {$\lambda^{\star}_{ij}$} (lp |- draw.north);
  \draw[->] (draw.west) -| node[lbl, pos=0.5, above] {$(i,j)$} ([xshift=1.5mm]law.north);
  \draw[->] (pair.south) -- node[lbl, left, pos=0.62] {$K^{ij}_k(l),\,v^{ij}_k,\,\Tilde{v}^{ij}_k(l)$} ++(0,-0.95) -| ([xshift=-5mm]law.north);
\end{tikzpicture}
\caption{Offline synthesis and runtime execution of \eqref{eq:mixtureControl}.}
\label{fig:pipeline}
\vspace{-10pt}
\end{figure}

Fig.~\ref{fig:pipeline} summarizes the synthesis and  runtime.
Executing \eqref{eq:mixtureControl} requires two indices at each step.
The pair $(i,j)$ is latent, so we draw it from $\gamma_{ij}(\cdot)$ via \eqref{eq:initRule} or \eqref{eq:stepRule}, while we estimate the disturbance component $J_k$ from the realized disturbance $\bfw_k$ by an assignment rule.
Any measurable map $\R^{p}\to[M_k]$ is a valid assignment rule.
We use
\begin{equation}\label{eq:argmax}
    J^{\mathrm{m}}_k = \underset{l\in[M_k]}{\arg\max}\ \alpha_{\bfw_k,l}
    \Ncal(\bfw_k;\mu_{\bfw_k,l},\Sigma_{\bfw_k,l}),\vspace{-4pt}
\end{equation}
from~\cite{celeux1992classification}.
Richer rules, such as sampling from the posterior $\Pr(J_k \mid \bfw_k)$~\cite{celeux1992classification} or deferring the assignment through a filter over $J_{0:k}$~\cite{alspach1972nonlinear}, are not considered here due to space limitations.
Section~\ref{sec:law} presumes we have $J_k$, whereas \eqref{eq:argmax} supplies $J^{\mathrm{m}}_k$, and the two disagree whenever the realized noise is more likely under another mixture component.
Let $\Delta K^{\mathrm{m}} := K^{ij}_k(J^{\mathrm{m}}_k) - K^{ij}_k(J_k)$. Such a mislabel adds to \eqref{eq:devRec} the error,
\begin{align}\label{eq:mislabel}
    \epsilon^{ij}_k := B_k\Big[\Delta K^{\mathrm{m}}\big(\bfx_k - \mu^{ij}_{\bfx_k}\big) + \Tilde{v}^{ij}_k(J^{\mathrm{m}}_k) - \Tilde{v}^{ij}_k(J_k)\Big].
\end{align}
The following result quantifies the effect of \eqref{eq:mislabel}.
\begin{prop}\label{prop:mislabel}
    Let the conditions of Proposition~\ref{prop:feasible} hold and fix $(i,j)$.
    Let $\pi^{\mathrm{m}}$ denote \eqref{eq:mixtureControl} with $J^{\mathrm{m}}_k$ in place of $J_k$, under any assignment rule, with state $\bfx^{\mathrm{m}}_k$ and input $\bfu^{\mathrm{m}}_k$, and let $\Delta\bfx_k := \bfx^{\mathrm{m}}_k - \bfx_k$ and $\Delta\bfu_k := \bfu^{\mathrm{m}}_k - \bfu_k$ be their deviations from the state and input under $J_k$.
    Then $\Delta\bfx_T = \sum_{k=0}^{T-1}\Phi_k\epsilon_k$, where $\epsilon_k$ is \eqref{eq:mislabel} evaluated at $\bfx^{\mathrm{m}}_k$ and $\Phi_k := \bar{A}_{T-1}(J_{T-1})\cdots\bar{A}_{k+1}(J_{k+1})$ with $\Phi_{T-1} := I_n$, and
    \begin{subequations}\label{eq:mismatch}
    \begin{align}
        \E[\bfx^{\mathrm{m}}_T \mid i,j] - \mu^f_j &= \textstyle\sum_{k=0}^{T-1}\E[\Phi_k]\,\E[\epsilon_k], \label{eq:mismatchMean}\\
        \Cov(\bfx^{\mathrm{m}}_T \mid i,j) - \Sigma^f_j &= \Cov(\Delta\bfx_T) + \Gamma + \Gamma^{\intercal}, \label{eq:mismatchCov}\\
        \big|\Tilde{C}(\pi^{\mathrm{m}})^{1/2} - \Tilde{C}(\pi)^{1/2}\big| &\leq D^{1/2}, \label{eq:mismatchCost}
    \end{align}
    \end{subequations}
    where $\Gamma$ is the cross-covariance of $\bfx_T$ and $\Delta\bfx_T$, and $D$ is \eqref{eq:cost}, with $x_{d,k} = 0$, evaluated at $\Delta\bfx_k$ and $\Delta\bfu_k$.
\end{prop}
\begin{proof}
    By \eqref{eq:mislabel}, the deviation $\bfx^{\mathrm{m}}_k - \mu_{\bfx_k}$ obeys \eqref{eq:devRec} with $\epsilon_k$ added to its right side.
    For the same $\bfx_0$ and noise, subtracting \eqref{eq:devRec} under $J_k$ gives $\Delta\bfx_{k+1} = \bar{A}_k(J_k)\Delta\bfx_k + \epsilon_k$ with $\Delta\bfx_0 = 0$, hence $\Delta\bfx_T = \sum_{k}\Phi_k\epsilon_k$.
    Since $\epsilon_k$ is a function of $\bfx_0$ and of $(J_t,\bfw_t)$ for $t\leq k$, Assumption~\ref{as:noise} makes it independent of $\Phi_k$, so $\E[\Phi_k\epsilon_k] = \E[\Phi_k]\E[\epsilon_k]$.
    By Proposition~\ref{prop:feasible}, $\bfx_T$ has mean $\mu^f_j$ and covariance $\Sigma^f_j$, so the mean and the covariance of $\bfx^{\mathrm{m}}_T = \bfx_T + \Delta\bfx_T$ are \eqref{eq:mismatchMean} and \eqref{eq:mismatchCov}.
    Finally, \eqref{eq:mismatchCost} is the triangle inequality for the norm whose square is \eqref{eq:cost} with $x_{d,k} = 0$, applied to the sum of $(\bfx_k - x_{d,k},\bfu_k)$ and $(\Delta\bfx_k,\Delta\bfu_k)$.
\end{proof}

\section{Terminal Covariance Reachability under Mixture Process Noise Preview}\label{sec:reach}

The remaining inputs cannot cancel noise that enters outside the subspace they reach.
This puts a floor under the terminal covariance, and limits when \eqref{eq:SDPcovMixture} is feasible.
Since \eqref{eq:SDPcovMixture} decouples across pairs, we fix $(i,j)$, and for clarity take $(A_k,B_k,G_k)\equiv(A,B,G)$ with a time-invariant mixture, $M_k\equiv M$ and $(\alpha_{\bfw_k,l},\mu_{\bfw_k,l},\Sigma_{\bfw_k,l})\equiv(\alpha_{\bfw,l},\mu_{\bfw,l},\Sigma_{\bfw,l})$, and write $\mathrm{Cov}(\bfw)$ for the covariance of $\bfw_k$, which no longer depends on $k$.
Let $\scriptR_q := \range([B,\ AB,\ \dots,\ A^{q-1}B])$ be the subspace of $\bfx_T$ that the last $q$ inputs can reach, $\scriptR_0 := \{0\}$, $Z_q$ an orthonormal basis of $\scriptR_q^{\perp}$, and $\nu := \min\{\, q \in \mathbb{N} : \scriptR_q = \mathbb{R}^n \,\}$ the controllability index, well defined with $\nu \leq \min(n,T)$ by Assumption~\ref{as:ctrb}.
\begin{prop}\label{prop:reach}
Let $\bar\Sigma_{\bfw} := \sum_{l\in[M]}\alpha_{\bfw,l}\Sigma_{\bfw,l}$.
If
\vspace{-8pt}\setlength{\multlinegap}{0pt}
\begin{multline}\label{eq:F3}
    Z_q^{\intercal}\Sigma^f_jZ_q \succeq Z_q^{\intercal}
      \Big[\sum_{s=0}^{q-1}A^{s}G\,\mathrm{Cov}(\bfw)\,G^{\intercal}(A^{s})^{\intercal} + P_q\Big]Z_q
\end{multline}
where $P_q=A^{q}G\bar\Sigma_{\bfw}G^{\intercal}(A^{q})^{\intercal}$, fails for some $q \in \Z_{0:\nu-1}$, then \eqref{eq:SDPcovMixture} is infeasible.
\end{prop}
\begin{proof}
By contraposition, suppose \eqref{eq:SDPcovMixture} is feasible, with \eqref{eq:lmiHat} tight without loss, and fix $q < \nu \leq T$.
Iterating \eqref{eq:sys} and \eqref{eq:meanRec} over the last $q+1$ steps, projecting on $\scriptR_q^{\perp}$, where $Z_q^{\intercal}A^{s}B = 0$ for $s < q$ since $\range(A^{s}B) \subseteq \scriptR_{s+1} \subseteq \scriptR_q$, and splitting the noise at $s = q$ by \eqref{eq:offset} gives
\begin{align}
Z_q^{\intercal}\big(\bfx_T - \mu^{ij}_{\bfx_T}\big) &= \xi + Z_q^{\intercal}A^{q}G\big(\bfw_{T-1-q} - \mu_{\bfw,J_{T-1-q}}\big) \nonumber\\
&\quad + \sum_{s=0}^{q-1} Z_q^{\intercal}A^{s}G\big(\bfw_{T-1-s} - \bar\mu_{\bfw}\big), \label{eq:proj}
\end{align}
where $\xi := Z_q^{\intercal}\big[A^{q+1}(\bfx_{T-1-q} - \mu^{ij}_{\bfx_{T-1-q}}) + A^{q}B(\bfu^{ij}_{T-1-q} - v^{ij}_{T-1-q}) + A^{q}d(J_{T-1-q})\big]$ is a function of $(\bfx_{T-1-q}, J_{T-1-q})$ under \eqref{eq:mixtureControl}.
By Definition~\ref{def:gmm} and Assumption~\ref{as:noise}, the other terms of \eqref{eq:proj} are zero mean given $(\bfx_{T-1-q}, J_{T-1-q})$ and mutually independent, and their second moments are the two terms of \eqref{eq:F3}.
Hence $Z_q^{\intercal}\Sigma^f_jZ_q$ equals the right side of \eqref{eq:F3} plus $\E[\xi\xi^{\intercal}] \succeq 0$.
\end{proof}
The $\nu$ inequalities of \eqref{eq:F3} evaluate a target $\Sigma^f_j$ before any instance of \eqref{eq:QPmean} or \eqref{eq:SDPcovMixture} is solved.
The noise entering at step $T-1-s$ reaches $\bfx_T$ through $A^{s}G$.
The $s$ later inputs act on it through the state and reach $\scriptR_s$.
Under \eqref{eq:mixtureControl} the input at the same step depends on it only through its label and reaches $\range(A^{s}B)$.
For $s<q$ both subspaces lie in $\scriptR_q$, so no input acts on $\scriptR_q^{\perp}$ and all of $\mathrm{Cov}(\bfw)$ enters the floor.
At $s=q$ the input at the same step reaches $\range(A^{q}B)$, which is not contained in $\scriptR_q$, and the offset $d(J_{T-1-q})$ drops out of the floor.
Preview therefore lowers the floor of \cite[Thm.~3]{liu2024reachability}.

Replacing $\bar\Sigma_{\bfw}$ by $\mathrm{Cov}(\bfw)$ in \eqref{eq:F3} recovers a floor which ignores the preview and rejects reachable targets.
Naively replacing $\mathrm{Cov}(\bfw)$ by $\bar\Sigma_{\bfw}$ gives \cite[Thm.~3]{liu2024reachability}, which admits targets no controller reaches.
Condition \eqref{eq:F3} sits between, and depends on how much of $G(\Cov(\bfw)-\bar\Sigma_{\bfw})G^\intercal
            = \sum_{l\in[M]}\alpha_{\bfw,l}\,d(l)d(l)^\intercal$ lies outside $\scriptR_{q}$.
If $d(l)\in\range(B)$ for every $l$, then $A^{s}d(l)\in\scriptR_q$ for $s < q$, and $\mathrm{Cov}(\bfw)$ may be replaced by $\bar\Sigma_{\bfw}$ in \eqref{eq:F3}.
The separation between component means then does not contribute to the floor.

\vspace{-4pt}
\section{Numerical Example}\label{sec:examples}
\vspace{-13pt}
\begin{figure}[h!]
    \centering
    \includegraphics[width=\linewidth]{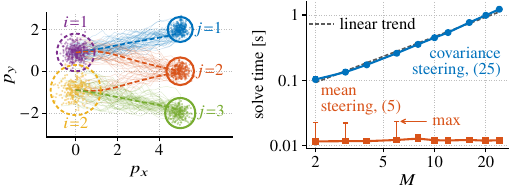}
    \vspace{-18pt}
\caption{GMM steering in position under preview (left) and solve times against the number of noise components $M$.}
    \label{fig:steerComputation}
\end{figure}
\vspace{-24pt}
\begin{figure}[ht!]
    \centering
    \includegraphics[width=\linewidth]{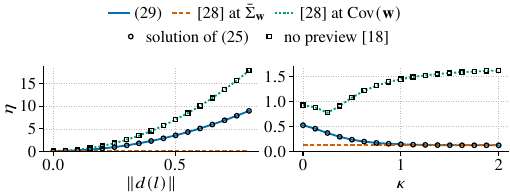}
    \vspace{-18pt}
\caption{Smallest $\eta$ for which $\eta\Sigma^f_1$ satisfies each condition, against the mean over $l$ of $\|d(l)\|$ in fixed directions outside $\range(B)$ (left) and against $\kappa$, which rotates $d(l)$ at fixed length into $\range(B)$ at $\kappa=2$ (right).}
    \label{fig:reach}
    \vspace{-4pt}
\end{figure}
\begin{figure}[ht!]
    \vspace{4pt}
    \centering
    \includegraphics[width=\linewidth]{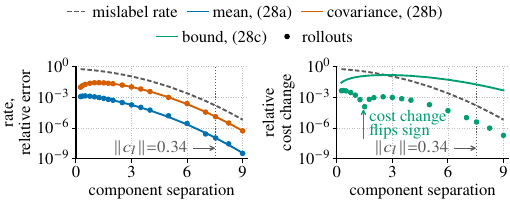}
    \vspace{-16pt}
\caption{Mislabel rate, \eqref{eq:mismatchMean}, and \eqref{eq:mismatchCov} (left), and relative change in cost with \eqref{eq:mismatchCost} (right), against the Mahalanobis separation of the disturbance mixture components.}
    \label{fig:sensitivity}
\end{figure}

Consider the 2D double integrator with $\bfx = [p_{x}\ p_{y}\ \dot p_{x}\ \dot p_{y}]^{\intercal}$, $\Delta T = 0.5$, $T = 20$, $G = I_{4}$, $Q_{k} = 0.1 I_{4}$, $R_{k} = I_{2}$, and $x_{d,k}$ the straight-line path from $\mu^{0}_{i}$ to $\mu^{f}_{j}$~\cite{sivaramakrishnan2022distribution}.
The initial GMM has $N_{0} = 2$ with $\alpha^{0} = [0.60,\ 0.40]^{\intercal}$ and $\mu^{0}_{1,2} = [0,\ \pm 0.9,\ 1,\ \pm 0.12]^{\intercal}$, and the terminal GMM has $N_{f} = 3$ with $\alpha^{f} = [0.30,\ 0.40,\ 0.30]^{\intercal}$ and $\mu^{f}_{j} = [5,\ \beta_{j},\ 0,\ 0.1\beta_{j}]^{\intercal}$ for $\beta_{j} \in \{2, 0, -2\}$.
The noise GMM has $M_{k} \equiv 3$ with $\alpha_{\bfw} = [0.35,\ 0.40,\ 0.25]^{\intercal}$ and $\mu_{\bfw,l} = [\Delta T\,c_l^{\intercal},\ \kappa\,c_l^{\intercal}]^{\intercal}$, where $\kappa = 0.2$ and $c_l = 0.34\,[\cos\theta_l,\ \sin\theta_l]^{\intercal}$ with $\theta_l \in \{90^{\circ}, 210^{\circ}, 330^{\circ}\}$.
All covariances are $\Sigma = (\sigma_{\cdot})^{2}\blkdiag(I_{2}, \tfrac{1}{4}I_{2})$, with $\sigma_{0} = (0.3, 0.4)$, $\sigma_{f} = (0.2, 0.22, 0.24)$, and $\sigma_{\bfw} = 0.05$.
We solve \eqref{eq:QPmean} and \eqref{eq:SDPcovMixture} with CVXPY and Clarabel at a tolerance of $10^{-9}$ and \eqref{eq:gmmLP} with SciPy, on an M4 Max CPU.

Fig.~\ref{fig:steerComputation} uses law \eqref{eq:initRule}. The median solve times over $10$ runs are $12$~ms per pair for \eqref{eq:QPmean}, $0.14$~s per pair for \eqref{eq:SDPcovMixture}, and $0.5$~ms for \eqref{eq:gmmLP}.
Propagating the recovered gains and feedforwards through \eqref{eq:meanRec} and \eqref{eq:covRec} reproduces $\mu^{f}_{j}$ and $\Sigma^{f}_{j}$ to relative errors of $9.72\times10^{-13}$ and $9.31\times10^{-13}$, and \eqref{eq:lmiHat} is tight to $5.90\times10^{-11}$ in Frobenius norm.

The controllability index is $2$ and \eqref{eq:F3} binds at $q = 1$, where $\scriptR_1 = \range(B)$, and Fig.~\ref{fig:reach} shows that \eqref{eq:F3} tracks the attainable floor.
Steering without preview, i.e., \cite{rapakoulias2026mixture} with the noise replaced by a Gaussian of the same mean and covariance, costs $34.9\%$ more for the same targets, and its smallest reachable multiple of $\Sigma^f_j$ is $2.08$ times larger, a ratio that reaches $12.9$ in Fig.~\ref{fig:reach}.

The rule \eqref{eq:argmax} supplies $J^{\mathrm{m}}_k$ with a mislabel rate of $1.59\times10^{-4}$, so the mismatch \eqref{eq:mislabel} is non-zero.
Over $N_{\text{samples}} = 2\times10^{5}$ rollouts its $\Delta K^{\mathrm{m}}$ term stays below $10^{-6}$ while its deviation feedforward term reaches $1.93\times10^{-1}$.
Evaluating \eqref{eq:mismatchMean} and \eqref{eq:mismatchCov} under \eqref{eq:initRule} gives terminal errors of at most $1.09\times10^{-7}$ in the mean and $1.32\times10^{-5}$ in the covariance, relative to $\|\mu^f_j\|$ and $\|\Sigma^f_j\|_F$.
Fig.~\ref{fig:sensitivity} scales $c_l$ to vary the separation of the components.
The covariance error peaks at $2.72\times10^{-2}$ at a separation of $1.25$, where $39\%$ of the labels are wrong, and the relative change in cost is at most $4.72\times10^{-3}$ against a bound from \eqref{eq:mismatchCost} of at most $1.55\times10^{-1}$.

\vspace{-3pt}
\section{Conclusion}

This letter presents a convex solution to the GMM distribution steering problem under previewed GMM disturbances.
Indexing the policy on disturbance mixture components yields a QP and an SDP per boundary pair, with an LP coupling the overall transport cost.
Since the mixture component is estimated at runtime from realizations of the disturbance, Proposition~\ref{prop:mislabel} gives the errors in the terminal mean and covariance and a bound on the change in cost.
Proposition~\ref{prop:reach} bounds attainable terminal covariances under preview.

\vspace{-3pt}
\section*{Acknowledgments}

Anthropic's Claude Opus 5.5 \cite{anthropic2026claude} was used to revise the TikZ for Fig.~\ref{fig:overview} and Fig.~\ref{fig:pipeline}, locate and suggest candidate references, correct bibliographic entries to ISO~4 abbreviations, revise prose for clarity and length, and debug code for the numerical example.
The authors reviewed and edited all content and take full responsibility for the final work.

\vspace{-4pt}
\bibliographystyle{IEEEtran}
\bibliography{IEEEabrv,shortIEEE,refs}

\end{document}